\documentclass[3p,times,twocolumn]{elsarticle}

\usepackage{ecrc}
\usepackage{verbatim}

\volume{00}

\firstpage{1}

\journalname{Nuclear and Particle Physics Proceedings}

\runauth{}

\jid{nppp}

\jnltitlelogo{Nuclear and Particle Physics Proceedings}

\usepackage{amssymb}

\usepackage[figuresright]{rotating}

\begin{document}

\begin{frontmatter}



\dochead{}

\title{GALPROP cosmic-ray propagation code: recent results and updates}


\author[label1]{Elena Orlando}
\ead{eorlando@stanford.edu}
\author[label2]{Gudlaugur Johannesson}
\author[label1]{Igor V. Moskalenko} 
\author[label1]{Troy A. Porter}
\author[label3]{Andrew Strong}

\address[label1]{Hansen Experimental Physics Laboratory, Stanford University, Stanford, CA 94305, USA ; Kavli Institute for Particle Astrophysics and Cosmology, Stanford University, Stanford, CA 94305, USA}
\address[label2]{Science Institute, University of Iceland, Dunhaga 3, IS-107 Reykjavik, Iceland}
\address[label3]{Max-Planck-Institut f\"ur extraterrestrische Physik, Postfach 1312, D-85741 Garching, Germany}

\begin{abstract}

Information on cosmic-ray (CR) composition comes from direct CR measurements while their distribution in the Galaxy is evaluated from observations of their associated diffuse emission in the range from radio to gamma rays. Even though the main interaction processes are identified, more and more precise observations provide an opportunity to study more subtle effects and pose a challenge to the propagation models.

GALPROP is a sophisticated CR propagation code that is being developed for about 20 years. It provides a unified framework for interpretations of data from many different types of experiments. It is used for a description of direct CR measurements and associated interstellar emissions (radio to gamma rays), thereby providing important information about CR injection and propagation in the interstellar medium. By accounting for all relevant observables at a time, the GALPROP code brings together theoretical predictions, interpretation of the most recent observations, and helps to reveal the signatures of new phenomena.

In this paper we review latest applications of GALPROP and address ongoing and near future improvements. 
We are discussing effects of different propagation models, and of the transition from cylindrically symmetrical models to a proper 3D description of the components of the interstellar medium and the source distribution.
\end{abstract}


\end{frontmatter}


\section{Introduction}\label{Introduction}

Understanding the propagation of CRs in the interstellar medium and in the heliosphere is equally important. One single point in the Galaxy where we can reconcile the direct and indirect measurements of CRs is located deep inside of the heliosphere. Even though the size of the heliosphere is incomparable to the size of the Galaxy, it "affects" the whole Galaxy. In fact the most precise isotopic measurements are made at low energies and the information gained from them is extended to all energies up to the knee and to the whole Galaxy. Fortunately for the life on Earth, the heliosphere is blocking about 75\% of Galactic CRs, but simultaneously it is blocking scientists from getting access to the interstellar fluxes of CRs. Getting through this bottle neck was almost impossible until recently.

With the technology and software of early 1970s, Voyager 1 spacecraft is still beaming back everything it sees or measures during its $\sim$40 years in space. Since the end of August of 2012, it is finally leaving the heliosphere. The measurements made by the spacecraft since then put an anchor on the models of heliospheric propagation now on. This allows the existing bottle neck to widen significantly with the hope that it will drop completely very soon. 

In this paper we review recent results of our work with our colleagues completed during the latest years that was aimed at making both the interstellar propagation and solar modulation as easy and accessible as the famous leaky-box/force-field combination just a decade ago, but at a much higher level of sophistication. The latter is a must to take full advantage of the high precision of nowadays experiments. Then, we discuss some new scenarios for propagation models.

\section{The GALPROP code}\label{GALPROP}

The GALPROP\footnote{http://galprop.stanford.edu/} code stays around for about 20 years now. During these years, it became a "standard model" in the astrophysics of CRs. Even though it does not describe the CR transport from the first principles, the diffusion approximation works well enough and allows many ideas and hypotheses to be easily tested against an array of all kinds of relevant data \cite{Strong2007,Vla}.

GALPROP solves the transport equation  for  a  given  CR  source  distribution  and  boundary
conditions \cite{SM98}, for all CR species ($Z\le28$). It takes into account energy losses, diffusion, stochastic reacceleration, convection, fragmentation, radiative decay, and constraints coming from multi-wavelength observations \cite{Strong2007}.  Secondary production in
collisions of CR particles with interstellar gas and the following decay of radiative isotopes are
included.  

The propagation equation is solved numerically on a user-defined spatial grid in 2D or in 3D plus
a momentum grid. The iterations proceed until a steady-state
solution is obtained, starting with the heaviest primary
nucleus, and then electrons, positrons, and antiprotons
are also computed. GALPROP calculates the three components of the diffuse emission, neutral pion decay, bremsstrahlung, and inverse Compton, for a user-defined CR source distribution and injection spectra.

GALPROP calculates CR spectra and abundances, diffuse gamma-ray and synchrotron
emission, as well as the polarization of the latter \cite{O&S2013}. In the past,
GALPROP models were used for interpretation of the diffuse gamma-ray emission observed by two CGRO instruments EGRET and COMPTEL \cite{Moskalenko2000,SMR00,SMR04}, and by a hard X-ray -- soft gamma-ray mission INTEGRAL \cite{Porter}. Since the launch of the Fermi observatory, it is officially used by the Fermi-LAT collaboration for interpreting the diffuse emission data \cite{diffuse,diffuse2}. The GALPROP-based calculations of the Galactic radio-emission were used for the analysis of the WMAP, and Planck data \cite{O&S2013,Strong2011,Orlando2015}. Recently, GALPROP formalism for calculation of the interstellar synchrotron emission has been improved and extended to include synchrotron polarization, free-free emission and absorption \cite{O&S2013}.

\section{Interpretation of CR measurements}\label{results}

\subsection{Voyager 1 measurements}

\begin{figure}
\includegraphics[width=18pc]{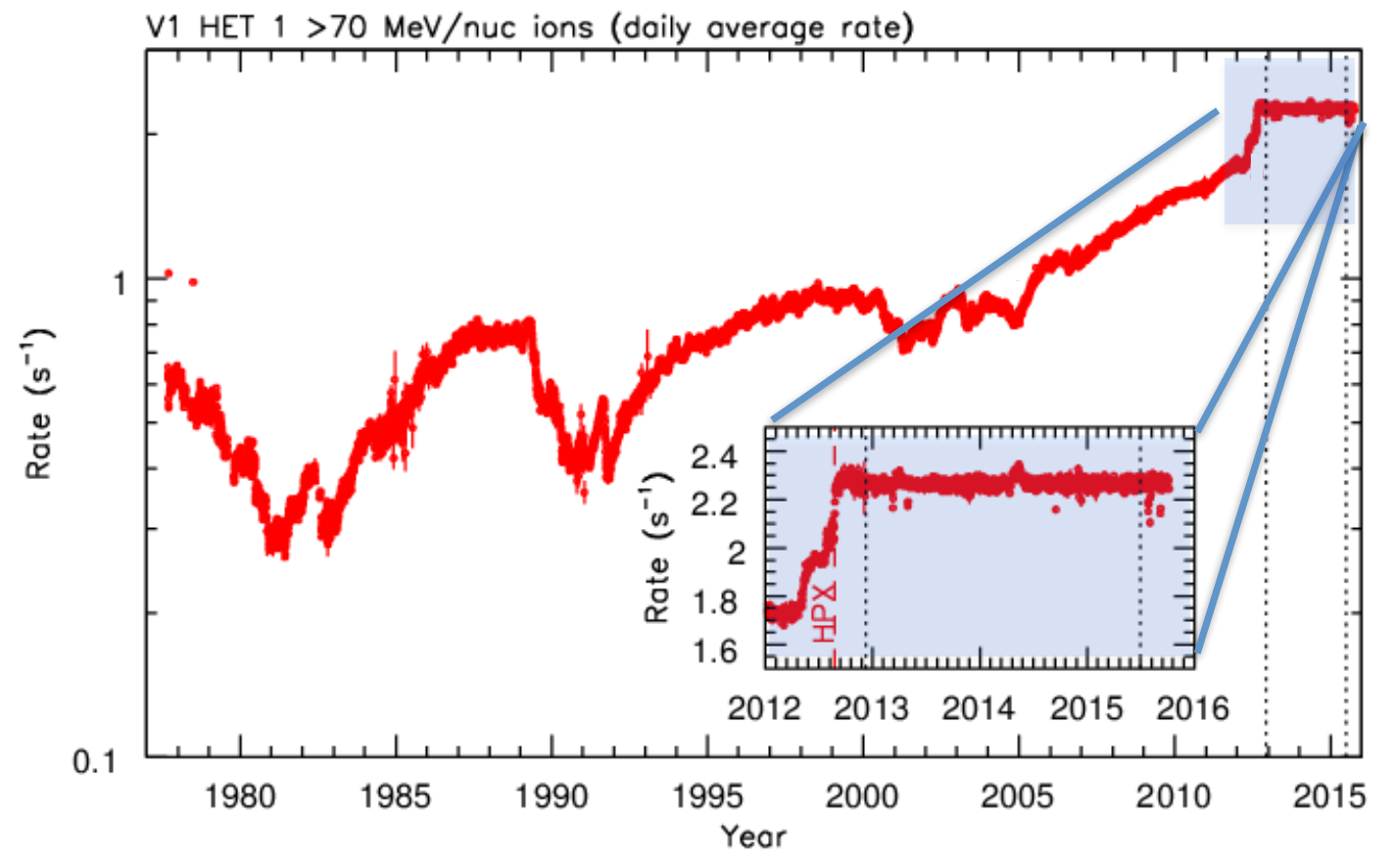}
\caption{Voyager 1 counting rate dominated by protons $\ge$70 MeV from 1977 through 2015 \cite{Voyager}. The inset shows the time period since 2012. The vertical dotted lines show the period selected for the analysis in the paper.}
\label{Fig1}
\end{figure}

Since the end of August of 2012, Voyager 1 observes \cite{Voyager} a steady flux of Galactic CRs down to 3 MeV/nucleon for nuclei  and to 2.7 MeV for electrons (Fig.\ref{Fig1}), which is independent on the solar activity. This is a strong indication of the instruments measuring the true CR spectra in the interstellar space.

GALPROP plain diffusion and diffusive-reacceleration models with standard propagation parameters show good agreement with Voyager 1 measurements of CR species from H to Ni in the energy range 10--500 MeV/nucleon \cite{Voyager}. Diffusive-reacceleration models are statistically favored with high significance. 

What is the reason of such a good agreement with data? Apparently, the most likely reason is the absence of a recent source of low energy CR hadrons in the solar system neighborhood. In the absence of such a CR source, the shape of the spectra of CR species at low energies is driven by the energy losses, mostly due to the ionization, which are properly accounted for by the GALPROP code. The rollover observed in the energy spectra at low energies from a power law at high energies, clearly visible in H and He spectra, is another indication that Voyager~1 is not in the vicinity of a recent source of CRs.

Among all-secondary Li, Be, and B nuclei, only B measurements have a couple of low energy data points below 30 MeV/nucleon that show an excess over the model predictions. Measurements of the other species demonstrate a good  agreement with model predictions. The inclusion of the Local Bubble and variations of the interstellar gas distribution does not cure the problem. Therefore, the observed excess is most likely due to the inaccurate description of B production at very low energies, or due to the limited statistics.

The ionization rate of atomic hydrogen inferred from the model fitting to the Voyager 1 data appears to be in the range $(1.51 - 1.64) \times 10^{-17}$ s$^{-1}$, which is a factor of $\sim$10 lower than the rate inferred using astrochemistry methods for diffuse interstellar clouds. This may be an indication that the local \emph{interarm} low-energy CR density is lower than the CR density in the Galactic arms where most of the neutral gas and CR sources are located. The energy density of CRs in the interstellar medium is about $0.83 - 1.02$ eV cm$^{-3}$.

\subsection{Bayesian analysis of CR propagation}

The Bayesian method provides not only a global best-fit point, but also statistically well-defined
uncertainties on model parameters. While very detailed numerical models of CR propagation exist,
a quantitative statistical analysis of such models has been so far hampered by the large computational
effort required. Statistical analyses have been carried out before using semi-analytical
models, but the evaluation of the results obtained from such models is difficult, as they necessarily
suffer from many simplifying assumptions.

We recently demonstrated \cite{Trotta} that a fully Bayesian parameter estimation can
be carried out with the GALPROP code, despite the heavy computational demands. In a recent work 
\cite{Galbayes} we performed a Bayesian search of the main GALPROP parameters, using the MultiNest nested sampling algorithm, enhanced by the BAMBI neural network machine-learning package. This is the first study to separate out low-mass isotopes ($p, \bar p$, and He) from the usual light elements (Be, B, C, N, and O). We found that the propagation parameters that best-fit $p, \bar p$, and He data are significantly different from those that fit light elements, including the B/C and $^{10}$Be/$^9$Be secondary-to-primary ratios normally used to calibrate propagation parameters. 

\begin{figure} 
\includegraphics[width=18pc]{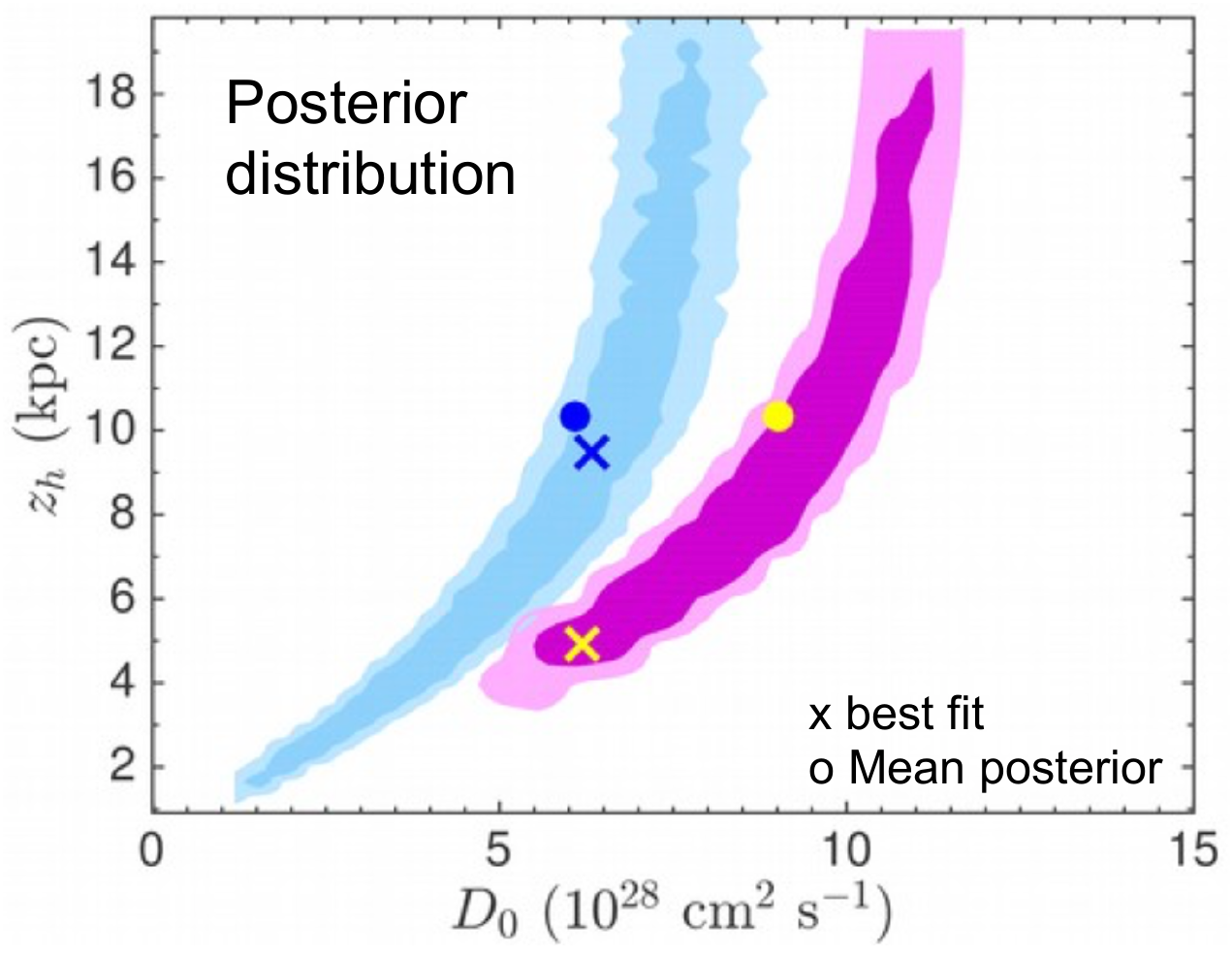}
\caption{Posterior distributions, mean and best-fit, with 1 and 2-sigma credible intervals for the $p, \bar p$, and He scan (blue), and for the light element scan (magenta). Adapted from \cite{Galbayes}.}
\label{Fig2}
\end{figure}

Fig.\ \ref{Fig2} shows the 2D correlation plot of the halo size vs.\ the normalization of the diffusion coefficient for both sets of data. The two plots clearly do not overlap suggesting that each set of species is probing a very different interstellar medium, and that the standard approach of calibrating propagation parameters for all species using B/C could lead to incorrect results.

\subsection{Solar modulation with HelMod}

Accounting for heliospheric propagation is absolutely essential
for proper treatment of CR propagation in the Galaxy, but was a considerable challenge until recently. These last 100 AU are so important because they provide a link between the predictions of the interstellar propagation models with the location where most of direct CR measurements are made. Even though, the heliospheric modulation affects only particles with small to medium energies below 30--50 GeV, this range includes the sub-GeV energies where the most precise measurements of CR isotopic composition are made. These low energy data are used to derive the parameters of interstellar propagation that are then extrapolated onto the whole Galaxy and energies up to the multi-TeV region.

We are recently teamed with a group of experts developing the HelMod\footnote{http://www.helmod.org/}  code that computes the transport of Galactic CRs through the heliosphere down to the Earth. The code is solving a time-dependent Parker equation. The solution is obtained using a 2D Monte Carlo approach and includes diffusion, convection, energy losses, and a general description of the symmetric and antisymmetric parts of the diffusion tensor, thus, properly addressing the charge-sign dependent drift effects. The model is being tuned to fit the Ulysses observations
outside the ecliptic plane (up to $\pm80^\circ$) at several distances from the Earth and the spectra observed near the Earth for both, high and low solar activity levels. A stand-alone Python module, fully compatible with GALPROP, has being developed to properly calculate the solar modulation. The local interstellar spectra of $p$ and He derived from a combined GALPROP/HelMod fit is described in \cite{Boschini}. A propagation model with acceleration and convection was found to give the best agreement with proton, helium, and antiproton data by AMS-02, BESS, PAMELA and {\it Voyager 1} from 1997 to 2015.

\section{CR-induced interstellar emission from radio to gamma rays}


\subsection{Radio and microwave synchrotron studies}

In 2011 we proposed \cite{Strong2011} a way to probe the local interstellar electron spectrum comparing GALPROP models of the synchrotron emission with a collection of radio surveys and WMAP. We found that in order to reproduce the synchrotron data, the local interstellar electron spectrum should have a break at a few GeV where the value of the spectral index changes from $<$2 to 3. The change of the spectral index was found to be independent of propagation models. Besides, we found that the injection spectrum below a few GeV should be harder than 1.6. Plain diffusion models described the data reasonably well, while it was more challenging for reaccleleration models at low frequencies. There, the latter were producing too many secondaries that overshot the synchrotron emission data.

More recently \cite{O&S2013}  the total and polarized synchrotron emission were investigated for the first time in the context of physical models of CR propagation and with 3D magnetic field models. Model predictions were compared with radio surveys from 22 MHz to 2.3 GHz and WMAP data at 23 GHz. After tuning the models to the Fermi all-electron measurements, we found that the all-sky total intensity and polarization maps were reasonably reproduced if an anisotropic component of the magnetic field is included; its intensity should be comparable to the regular component derived from the rotation measure. This also required a flat CR distribution in the outer Galaxy and an increased size of the halo. \\

The GALPROP model has been used also in recent studies within the Planck Collaboration.
At frequencies detected by Planck, the microwave sky is a superposition of different Galactic emission foregrounds that are very hard to disentangle. The best synchrotron spectral model from \cite{O&S2013} was used for the component separation and generation of the Planck maps officially released \cite{Planck_component}. In this work, where a detailed investigation of the low frequency foreground maps was preformed, some regions of the sky showed loops and spurs.
In the synchrotron polarization maps, we found that the spectral index of Loop I is harder than the local spectrum and the spectrum of a control loop. Consequently, this suggests that the CR electrons in Loop I have a harder spectrum as well.  Alternatively, it could suggest that the structure observed in the polarization map may be connected with the Fermi Bubble.  However, no significant variations of the spectrum across the bubbles were found in the analysis made in \cite{bubbles}. Besides, while the bubbles are projected as emanating from the Galactic center, Loop I is not and there is no obvious counterpart in the Southern hemisphere. This suggests that Loop I and Fermi Bubbles are two different structures at different distances \cite{Planck_component}.

A recent work \cite{Planck_bfield} investigated different Galactic magnetic field models taken from the literature, using a representative CR distribution from \cite{O&S2013}. Intensities of the polarized and random component of the Galactic magnetic field models were updated to the values that better reproduce the Planck maps.

Recently the latest AMS-02 all-electron data were used to improve the model predictions for the synchrotron emission \cite{Orlando2015}. A new propagation model reproduces the synchrotron spectral observations of the various radio surveys, and latest Planck maps and the 408MHz map reprocessed by \cite{408} (Orlando et al in preparation).  

\subsection{The gamma-ray sky}

A detailed study of the CR-induced diffuse emission from the whole Galaxy was performed on a grid of 128 propagation models \cite{diffuse2} using the Fermi-LAT data. Even though all models provide a good agreement with data, two issues came up. First, many model-dependent structures (e.g., Fermi bubbles, Loop I, outer Galaxy) showed up as excesses over the adopted model. 
Second, it was difficult to select a model that would provide the best description of the whole sky. This is likely due to fact that some models parameters are degenerate. 
However Fermi-LAT data show hints for a large propagation halo size, additional gas in the outer Galaxy, and/or a flat CR source distribution \cite{diffuse2}. New analyses of the all-sky diffuse emission are ongoing taking advantage of model improvements and more precise observations. \\
Further information on CRs in the Galaxy comes from gamma-ray observations of molecular clouds and comparisons with the large-scale models based on GAPROP, for example as done in \cite{Tibaldo}.

Two state-of-the-art analyses were recently published. In one of them, the GALPROP-based models were used to analyze the spectrum and morphology of the Fermi Bubbles, the huge 10-kpc-across structures emanating from the Galactic Center \cite{Su}. The analysis revealed that both lobes have well-defined edges and their spectra are surprisingly hard and very similar extending up to 200 GeV \cite{bubbles}. Another one was devoted to the analysis of the Fermi-LAT observations of the Galactic Center \cite{GC}. This analysis yielded a weak extended residual component peaked at the Galactic Center. The inverse Compton component was found to be dominant and enhanced in that region. If this is due to CR or interstellar radiation field is still an open question. Naturally, both analyses will be repeated with updated diffuse emission models.

\section{"Beyond" standard models}

This section investigates spatial differences in calculated gamma-ray and synchrotron maps of selected models with respect to standard models usually used in gamma-ray and radio studies. 

\begin{figure} [h!]
\includegraphics[width=18pc]{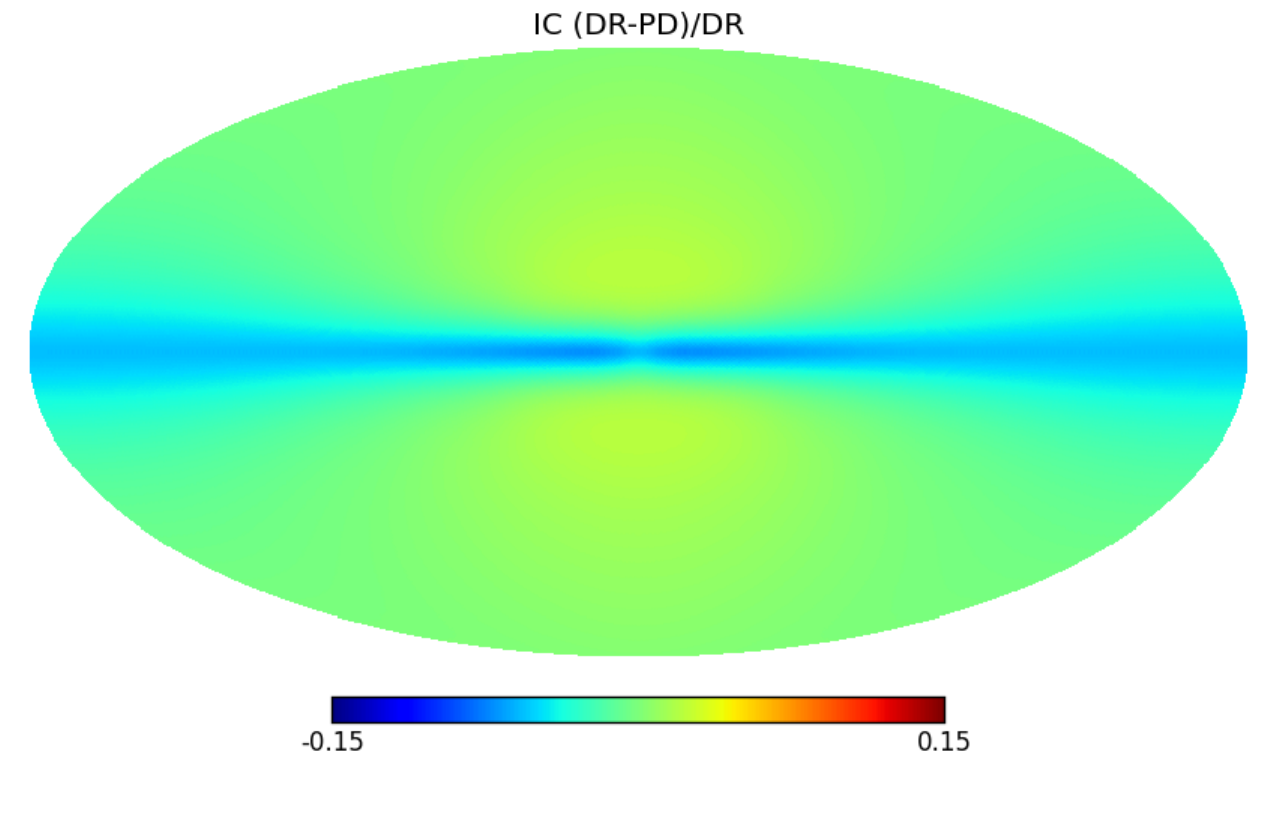}\\ 
\includegraphics[width=18pc]{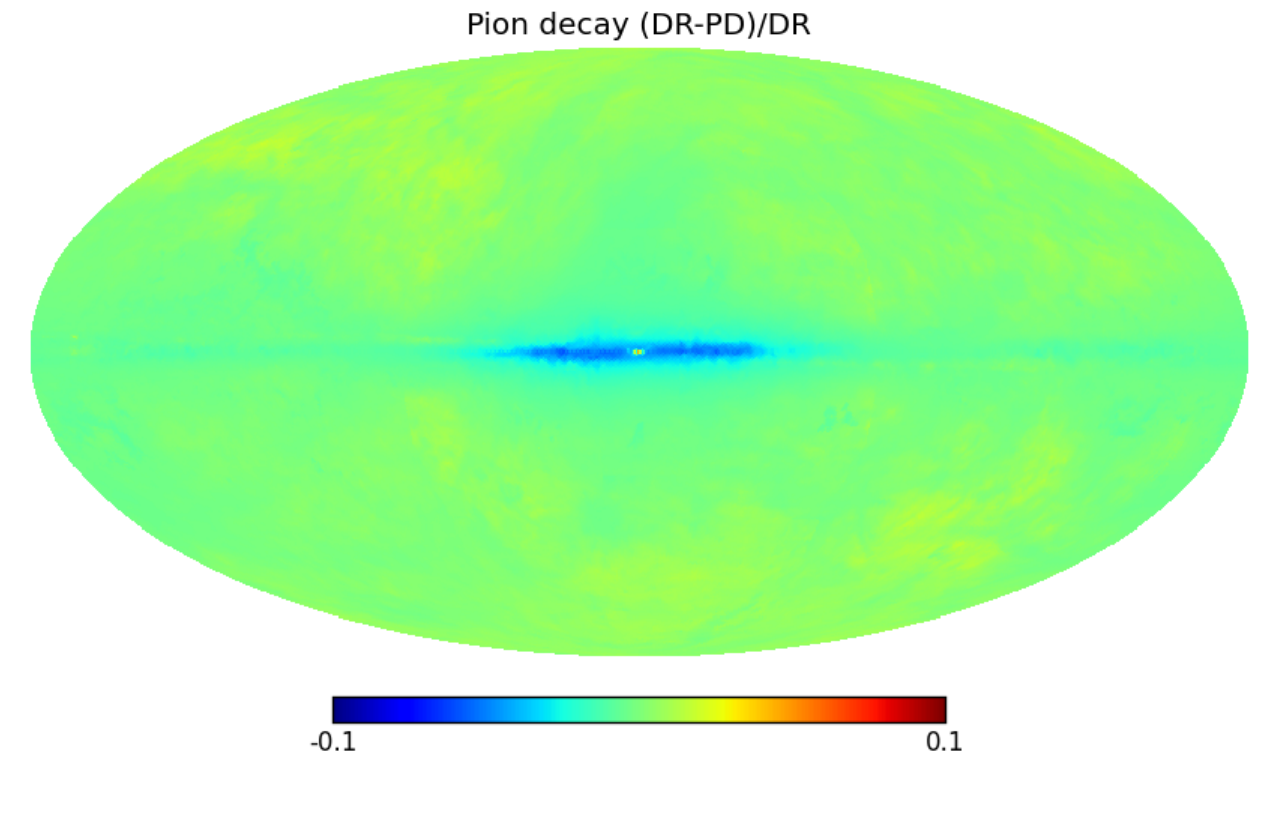}
\caption{Spatial differences of the Inverse Compton emission at 100~MeV (upper) and pion-decay emission at 1 GeV (lower) between a diffusion-reacceleration (DR) and a diffusion-only (PD) model. Skymaps show (DR-PD)/DR.} 
\label{Fig3}
\end{figure}

\begin{figure} [h!]
\includegraphics[width=18pc]{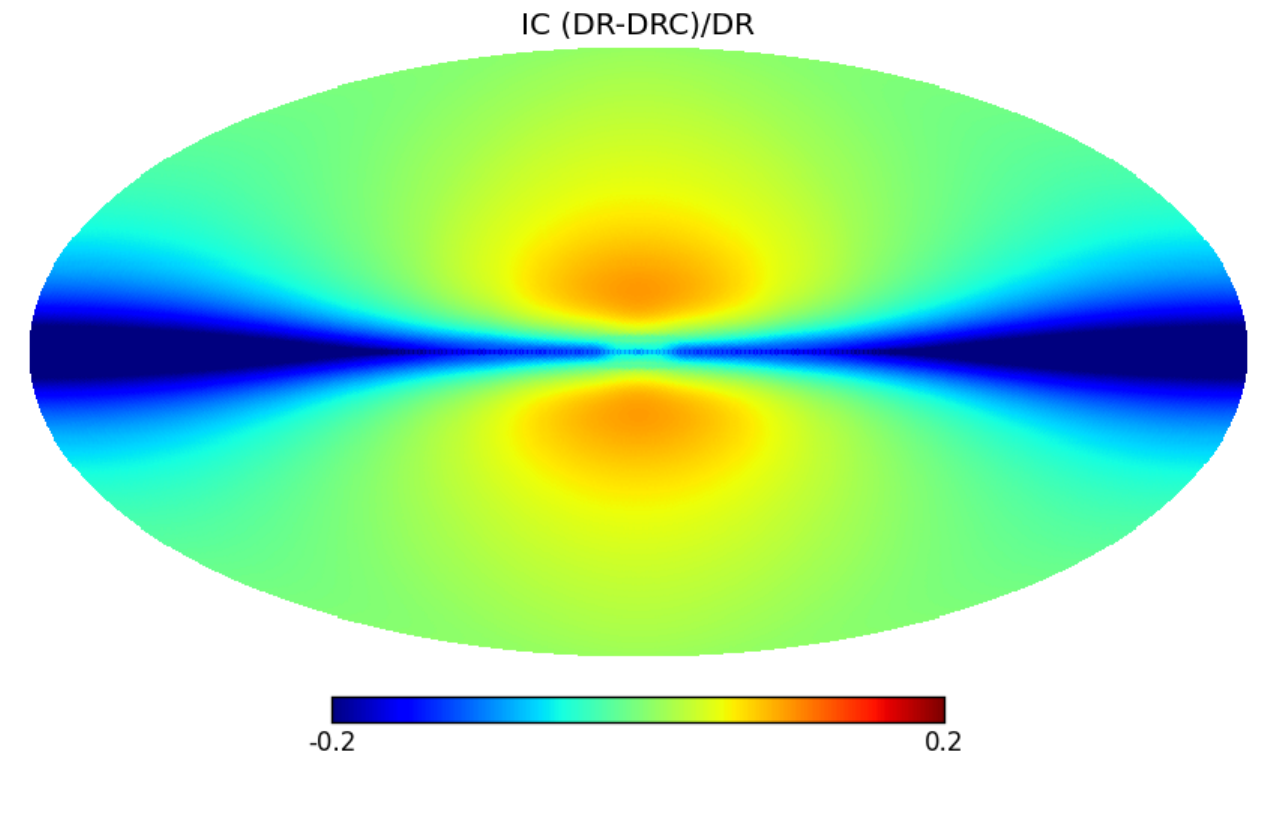}
\caption{Spatial differences of the Inverse Compton intensity at 100 MeV between a reacceleration model with no convection (DR) and a reacceleration model with convection (DRC). Skymaps shows (DR-DRC)/DR.} 
\label{Fig4}
\end{figure}

\begin{figure} [h!]
\includegraphics[width=18pc]{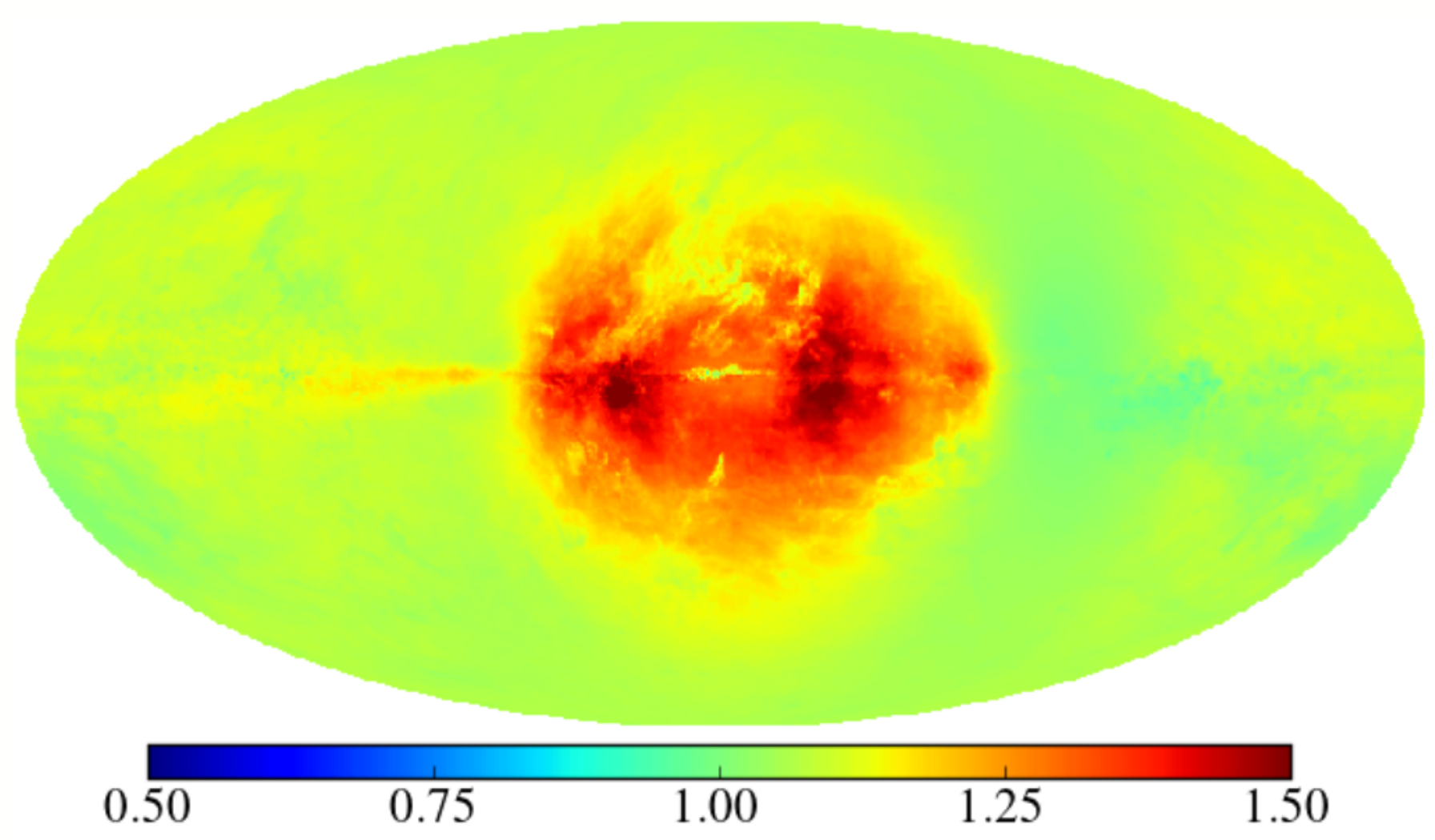}
\caption{Figure taken from \cite{galgas}. Ratio of the gamma-ray skymaps at 10 GeV calculated with GALPROP for two gas and CR source models: a model incorporating the spiral arm structure in the gas and CR source distributions vs. a standard model with cylindrical symmetry. The enhanced emission at the spiral arm tangents are visible in the map. Details can be found in \cite{galgas}.}
\label{Fig5}
\end{figure}

While reaceleration models have been extensively used in gamma-ray and radio data, pure diffusion models were tested with radio data only. As an example Figure \ref{Fig3} shows the spatial differences of the inverse Compton component and of the pion-decay component between a standard reacceleration model and a pure diffusion model, using the same CR source distribution, gas, ISRF and B-field models. We see that inclusion of re-acceleration processes provides up to 15\% spatial variation in the inverse Compton emission in the plane, and more than 10\% in the pion-decay emission at the Galactic center. Since synchrotron emission is produced by electrons, like inverse Compton emission, the same 15\% variation is seen in the synchrotron maps. 

Propagation models that include convection have never been tested against radio and gamma-ray data. Figure \ref{Fig4} shows up to 30\% spatial variation in the inverse Compton emission at 100 MeV between diffusion-reacceleration-convection and diffusion-reacceleration-only models, for the same CR source distribution, gas, ISRF and B-field models.

Fig.\ref{Fig5} instead shows an example of the variation of the calculated gamma-ray emission including spiral arm structures in the gas and CR source distributions with respect to the cylindrically symmetrical models. The figure is taken from \cite{galgas}.

\section{Future developments}
 
We aim at a fully 3-dimensional model of the interstellar medium that is one of the major constituents of the GALPROP model \cite{Moskalenko2015}. This is a significant task that comprises such ISM components as the molecular (H$_2$), atomic (H {\sc i}), and ionized (H {\sc ii}) gas, interstellar radiation field, and regular and random components of the magnetic field. Such a model would also serve as a basis for other studies, such as the interstellar gas dynamics and chemistry, star formation, foreground model of synchrotron emission and its polarization relevant for CMB studies.

\section{Summary}

The results discussed in this paper show that we are at the beginning of a new very exiting era in astrophysics. New effects are continuously being discovered. 
Their proper interpretation is impossible without a reliable self-consistent framework as the one provided by GALPROP. We have also investigated the effects of spatial models, which have never been tested against gamma-ray data so far.

\vspace*{0.5cm} 
\footnotesize{{\bf Acknowledgment:}{
 E.O. acknowledges support via NASA Grant No. NNX16AF27G. GALPROP development is supported through NASA grants No.\ NNX13AC47G and NNX17AB48G.}}







\end{document}